\documentclass[11pt]{article}
\usepackage{moriond,epsfig}

\def\be{\begin{equation}}
\def\ee{\end{equation}}
\def\bea{\begin{eqnarray}}
\def\eea{\end{eqnarray}}

\begin{document}
\vspace*{4cm}
\title{B mesons and form factors}

\author{ F. JUGEAU }

\address{Laboratoire de Physique Th\'eorique,\\
Unit\'e Mixte de Recherche UMR 8627-CNRS,\\
Universit\'e Paris-Sud XI, B\^atiment 210,\\
91405 Orsay Cedex, France\\
E-mail: frederic.jugeau@th.u-psud.fr}

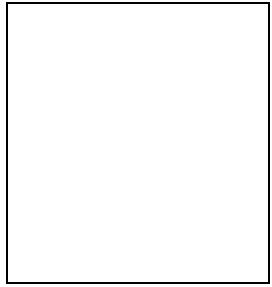
\maketitle\abstracts{In order to extract some information on the $CKM$ matrix element $|V_{cb}|$, we
have to determine form factors in $B$ decays. For this, we show that it is
relevant to consider the non-forward amplitude between the heavy-light $B$ and $D$
mesons within the Heavy Quark Effective Theory (HQET). This method provides us
crucial information on the shape of the elastic Isgur-Wise (IW) function
$\xi(w)$. As examples, one gets that the $n$-th derivative of $\xi(w)$ at $w=1$ can be
bounded by the $(n-1)$-th one and one obtains an absolute lower bound for the $n$-th
derivative $(-1)^{n}\xi^{(n)}(1)\geq\frac{(2n+1)!!}{2^{2n}}$. These bounds
should be taken into account in the parametrizations of $\xi(w)$ used to
extract $|V_{cb}|$. Besides, we have also obtained new information on
subleading functions at the order $O(1/m_{Q})$ where $m_{Q}$ is the heavy quark mass, namely $\xi_{3}(w)$ and $\overline{\Lambda}\xi(w)$.}

\section{Introduction}

In the leading order of the heavy quark expansion of QCD, Bjorken sum rule
(SR) \cite{Bjorken} relates the slope of the elastic IW function $\xi(w)$ to the
IW functions of transition $\tau_{1/2}^{(n)}(w)$ and $\tau_{3/2}^{(n)}(w)$
between the ground state $j^{P}=\frac{1}{2}^{-}$ and the
$j^{P}=\frac{1}{2}^{+}$, $\frac{3}{2}^{+}$ excited states at zero recoil where $w=1$ ($j$ is the total angular
momentum of the light cloud and $n$ is a radial quantum number). This
SR leads to the lower bound $-\xi'(1)=\rho^{2}\geq\frac{1}{4}$. Recently, a new
SR \cite{Uraltsev} was formulated by Uraltsev in the heavy quark limit involving also
$\tau_{1/2}^{(n)}(1)$ and $\tau_{3/2}^{(n)}(1)$ that implies, combined with Bjorken
SR, the much stronger lower bound $\rho^{2}\geq\frac{3}{4}$. A basic ingredient
in deriving this bound is the consideration of the non-forward amplitude
$B(v_{i})\rightarrow D^{(n)}(v')\rightarrow B(v_{f})$, allowing for general
$v_{i}$, $v_{f}$, $v'$ and where $B$ is a ground state meson. In order to
make a systematic study in the heavy quark limit of QCD, a manifestly
covariant formalism within Operator Product Expansion (OPE) has been developed
\cite{Oliver}, using the matrix representation for the whole tower of heavy meson
states \cite{Falk}. In this formalism, Uraltsev SR has been recovered plus a new class of SR
that allows to bound also higher derivatives of the IW function. Making
a natural physical assumption, this new class of SR implies the bound
$\sigma^{2}\geq\frac{5}{4}\rho^{2}$ where $\sigma^{2}$ is the curvature of the
IW function. Using this formalism including the whole tower of excited states
$j^{P}$, the rigorous bound $\sigma^{2}\geq\frac{5}{4}\rho^{2}$ has been
recovered plus generalizations that extend it to all the derivatives of the IW
function $\xi(w)$ at zero recoil, that is shown to be an alternate series in
power of $(w-1)$. After reviewing the corresponding sum rules in the heavy
quark limit of QCD, the object of the present paper is to extend the formalism to IW
functions at subleading order in $1/m_{Q}$.   

\section{Sum rules in the heavy quark limit of QCD}

Using the OPE and the trace formalism in the heavy quark limit, different
initial and final four-velocities $v_{i}$ and $v_{f}$, and heavy quark
currents $J_{1}=\overline{h}_{v'}^{(c)}\Gamma_{1}h_{v_{i}}^{(b)}$,
$J_{2}=\overline{h}_{v_{f}}^{(b)}\Gamma_{2}h_{v'}^{(c)}$ where $\Gamma_{1}$
and $\Gamma_{2}$ are arbitrary Dirac matrices, the following sum rule can be written \cite{Oliver}:
\begin{equation}
\begin{array}{c}
\{\Sigma_{D=P,V}\Sigma_{n}Tr[\overline{\mathcal{B}}_{f}(v_{f})\overline{\Gamma}_{2}\mathcal{D}^{(n)}(v')]\nonumber\\
\\
Tr[\overline{\mathcal{D}}^{(n)}(v')\Gamma_{1}\mathcal{B}_{i}(v_{i})]\xi^{(n)}(w_{i})\xi^{(n)}(w_{f})\nonumber\\
\\
+\;\;Other\;\;excited\;\;states\}\nonumber\\
\\
=-2\xi(w_{if})Tr[\overline{\mathcal{B}}_{f}(v_{f})\overline{\Gamma}_{2}P_{+}'\Gamma_{1}\mathcal{B}_{i}(v_{i})].\label{SR}
\end{array}
\end{equation}
In this formula, $v'$ is the intermediate charmed meson four-velocity,
$P_{+}'=\frac{1}{2}(1+v')$ is the positive energy projector on
the intermediate $c$ quark, $\xi(w_{if})$ is the elastic IW function
that appears because one assumes $v_{i}\neq v_{f}$. $\mathcal{B}_{i}$ and
$\mathcal{B}_{f}$ are the $4\times4$ matrices of the ground state $B$ or $B^{\ast}$
mesons, respectively the pseudoscalar and the vector of the doublet
$\frac{1}{2}^{-}$, while $\mathcal{D}^{(n)}$ stands for all possible ground state or excited
state $D$ mesons coupled to $B_{i}$ and $B_{f}$ through the currents. In
(\ref{SR}) we have made explicit the $j=\frac{1}{2}^{-}$ $D$ and $D^{\ast}$
mesons and their radial excitations of quantum number $n$. The variables
$w_{i}$, $w_{f}$ and $w_{if}$ are defined as $w_{i}=v_{i}.v'$,
$w_{f}=v_{f}.v'$, $w_{if}=v_{i}.v_{f}$ and vary within the domain \cite{Oliver}: 
\begin{equation}
\begin{array}{c}
w_{i}\geq1\hspace*{30pt},\hspace*{30pt}w_{f}\geq1\nonumber\\
\\
w_{i}w_{f}-\sqrt{(w_{i}^{2}-1)(w_{f}^{2}-1)}\leq w_{if}\leq w_{i}w_{f}+\sqrt{(w_{i}^{2}-1)(w_{f}^{2}-1)}.\label{domain}
\end{array}
\end{equation}

The general SR (\ref{SR}), obtained from the OPE, can be written in the
following compact way :
\begin{equation}
L_{Hadrons}(w_{i},w_{f},w_{if})=R_{OPE}(w_{i},w_{f},w_{if})\label{gsr}
\end{equation}
where the l.h.s. is the sum over the intermediate charmed D states, while the
r.h.s. is the OPE counterpart. Within the domain (\ref{domain}), one can derive
relatively to any of the variables $w_{i}$, $w_{f}$ and $w_{if}$ and obtain
different SR taking different limits to the frontier of the domain.

On the one hand, we can choose, as in ref. \cite{Oliver}, the $B$ pseudoscalar
meson $\mathcal{B}(v)=P_{+}(-\gamma_{5})$ as initial and final states and vector currents
projected along the $v_{i}$ and $v_{f}$ four-velocities :
\begin{equation}
J_{1}=\overline{h}_{v'}^{(c)}\not{v_{i}}h_{v_{i}}^{(b)}\hspace*{20pt},\hspace*{20pt}J_{2}=\overline{h}_{v_{f}}^{(b)}\not{v_{f}}h_{v'}^{(c)}.
\end{equation}
We then obtain SR (\ref{SR}) with the sum of all excited states $j^{P}$ in a compact form :
\begin{equation}
\begin{array}{c}
(w_{i}+1)(w_{f}+1)\Sigma_{l\geq0}^{}\frac{l+1}{2l+1}S_{l}(w_{i},w_{f},w_{if})\Sigma_{n}^{}\tau_{l+1/2}^{(l)(n)}(w_{i})\tau_{l+1/2}^{(l)(n)}(w_{f})\\
\\
+\Sigma_{l\geq1}^{}S_{l}(w_{i},w_{f},w_{if})\Sigma_{n}^{}\tau_{l-1/2}^{(l)(n)}(w_{i})\tau_{l-1/2}^{(l)(n)}(w_{f})\\
\\
=(1+w_{i}+w_{f}+w_{if})\xi(w_{if}).\label{premiere}
\end{array}
\end{equation}

On the other hand, choosing instead the axial currents, 
\begin{equation}
J_{1}=\overline{h}_{v'}^{(c)}\not{v_{i}}\gamma_{5}h_{v_{i}}^{(b)}\hspace*{20pt},\hspace*{20pt}J_{2}=\overline{h}_{v_{f}}^{(b)}\not{v_{f}}\gamma_{5}h_{v'}^{(c)}
\end{equation}
we get the following SR :
\begin{equation}
\begin{array}{c}
\Sigma_{l\geq0}^{}S_{l+1}(w_{i},w_{f},w_{if})\Sigma_{n}^{}\tau_{l+1/2}^{(l)(n)}(w_{i})\tau_{l+1/2}^{(l)(n)}(w_{f})\\
\\
+(w_{i}-1)(w_{f}-1)\sum_{l\geq1}^{}\frac{l}{2l-1}S_{l-1}(w_{i},w_{f},w_{if})\Sigma_{n}^{}\tau_{l-1/2}^{(l)(n)}(w_{i})\tau_{l-1/2}^{(l)(n)}(w_{f})\\
\\
=-(1-w_{i}-w_{f}+w_{if})\xi(w_{if}).\label{deuxieme}
\end{array}
\end{equation}
In (\ref{premiere}) and (\ref{deuxieme}), the IW functions $\tau_{l+1/2}^{(l)(n)}(w)$ and
$\tau_{l-1/2}^{(l)(n)}(w)$ have been defined in ref. \cite{Oliver} where $l$ and
$j=l\pm\frac{1}{2}$ are the orbital and the total angular momentum of the
light cloud and $S_{n}$ is given by
\begin{equation}
S_{n}(w_{i},w_{f},w_{if})=v_{i\nu_{1}}\ldots v_{i\nu_{n}}v_{f\mu_{1}}\ldots v_{f\mu_{n}}\Sigma_{\lambda}\epsilon'^{(\lambda)*\nu_{1}\ldots\nu_{n}}\epsilon'^{(\lambda)\mu_{1}\ldots\mu_{n}}
\end{equation}    
which can take the following expression \cite{Oliver}
\begin{equation}
S_{n}(w_{i},w_{f},w_{if})=\Sigma_{0\leq k\leq\frac{n}{2}}C_{n,k}(w_{i}^{2}-1)^{k}(w_{f}^{2}-1)^{k}(w_{i}w_{f}-w_{if})^{n-2k}
\end{equation}
with 
\begin{equation}
C_{n,k}=(-1)^{k}\frac{(n!)^{2}}{(2n)!}\frac{(2n-2k)!}{k!(n-k)!(n-2k)!}.
\end{equation}

On the one hand, making the sum of both equations (\ref{premiere}) and (\ref{deuxieme}), one obtains,
differentiating relatively to $w_{if}$ :
\begin{equation}
\xi^{(l)}(1)=\frac{1}{4}(-1)^{l}l!{\Large \{ }\frac{l+1}{2l+1}4\Sigma_{n}^{}{\large
  [}\tau_{l+1/2}^{(l)(n)}(1){\large ]}^{2}+\Sigma_{n}^{}{\large
  [}\tau_{l-1/2}^{(l-1)(n)}(1){\large ]}^{2}+\Sigma_{n}^{}{\large
  [}\tau_{l-1/2}^{(l)(n)}(1){\large ]}^{2}\}\hspace*{10pt}(l\geq0).\label{eq}
\end{equation}
This relation shows that $\xi(w)$ is an alternate series in power of
$(w-1)$ because of the factor $(-1)^{l}$ and reduces to Bjorken SR \cite{Bjorken} for $l=1$. 

On the other hand, differentiating (\ref{deuxieme}) relatively to $w_{if}$ and making
$w_{i}=w_{f}=w_{if}=1$, one has the simple formula :
\begin{equation}
\xi^{(l)}(1)=(-1)^{l}l!\Sigma_{n}^{}{\large
  [}\tau_{l+1/2}^{(l)(n)}(1){\large ]}^{2}\hspace*{20pt}(l\geq0).\label{ref2}
\end{equation}
Combining (\ref{eq}) and (\ref{ref2}), one obtains the relation among the IW
functions :
\begin{equation}
\frac{l}{2l+1}\Sigma_{n}[\tau_{l+1/2}^{(l)(n)}]^{2}-\frac{1}{4}\Sigma_{n}[\tau_{l-1/2}^{(l)(n)}]^{2}=\frac{1}{4}\Sigma_{n}[\tau_{l-1/2}^{(l-1)(n)}]^{2}
\end{equation}
that reduces to Uraltsev SR \cite{Uraltsev} for $l=1$ and generalizes it for
all $l$. Replacing now $\Sigma_{n}[\tau_{l+1/2}^{(l)(n)}]^{2}$ from the
expression (\ref{ref2}) into the generalization of Bjorken SR (\ref{eq}), one
obtains :
\begin{equation}
(-1)^{l}\xi^{(l)}(1)=\frac{1}{4}\frac{2l+1}{l}l!{\Large \{ }\Sigma_{n}^{}{\large
  [}\tau_{l-1/2}^{(l-1)(n)}(1){\large ]}^{2}+\Sigma_{n}^{}{\large
  [}\tau_{l-1/2}^{(l)(n)}(1){\large ]}^{2}{\Large \} }
\end{equation}
which implies the lower bounds :
\begin{equation}
\begin{array}{ccc}
(-1)^{l}\xi^{(l)}(1)&\geq&\frac{2l+1}{4}(-1)^{l-1}\xi^{(l-1)}(1)\\
\\
&\geq&\frac{(2l+1)!!}{2^{2l}}\label{ineq}
\end{array}
\end{equation}
that give, in particular, for the lower cases,
\begin{equation}
-\xi'(1)=\rho^{2}\geq\frac{3}{4}\hspace*{20pt},\hspace*{20pt}\xi''(1)\geq\frac{15}{16}.
\end{equation}

Considering systematically the derivatives of the SR (\ref{premiere}) and
(\ref{deuxieme}) relatively to $w_{i}$, $w_{f}$ and $w_{if}$ with the boundary
conditions $w_{i}=w_{f}=w_{if}=1$, one obtains a new SR :
\begin{equation}
\frac{4}{3}\rho^{2}+(\rho^{2})^{2}-\frac{5}{3}\sigma^{2}+\sum_{n\neq0}^{}|\xi^{(n)'}(1)|^{2}=0
\end{equation}
that in turn implies :
\begin{equation}
\sigma^{2}\geq\frac{1}{5}[4\rho^{2}+3(\rho^{2})^{2}]\label{truc}
\end{equation}
There is a simple intuitive argument to understand the last term in the bound
(\ref{truc}), namely the non-relativistic light quark $q$ interacting with a
heavy quark $Q$ through a potential. One can indeed prove that \cite{Oliver}
\begin{equation}
\sigma^{2}_{NR}\geq\frac{3}{5}(\rho^{2}_{NR})^{2}.\label{bb}
\end{equation}
Thus, the non-relativistic limit is a good guide-line to study the shape of
the IW function $\xi(w)$. We have recently generalized the bound (\ref{bb}) to
all the derivatives of $\xi_{NR}(w)$ \cite{Jugeau}. The method uses the positivity of
matrices of moments of the ground state wave function. We have
also shown that the method can be generalized to the real function $\xi(w)$ of QCD.

An interesting phenomenological remark is that the simple parametrization for
the IW function \cite{Morenas} 
\begin{equation}
\xi(w)={\Large (}\frac{2}{w+1}{\Large )}^{2\rho^{2}}\label{IW}
\end{equation}
satisfies the inequalities (\ref{ineq}) and (\ref{truc}) if
$\rho^{2}\geq\frac{3}{4}$. The result (\ref{ineq}), that shows that all derivatives at zero recoil are
large, should have important phenomenological implications for the empirical
fit needed for the extraction of $|V_{cb}|$ in the semileptonic exclusive
decay mode $B\rightarrow D^{\ast}l\nu$.

\section{Sum rules for subleading form factors}

We wish now to obtain new sum rules involving subleading quantities in
$1/m_{Q}$ where
$m_{Q}$ is the heavy quark mass \cite{Jugeau2}. For this, we can perturb the generic SR
(\ref{gsr}) by $1/m_{c}$ and $1/m_{b}$ terms. The perturbation of the r.h.s,
the OPE side, is parametrized by six new subleading IW functions concerning the ground state
$\frac{1}{2}^{-}$ and denoted by $L_{i}(w)$ ($i=1,\ldots,6$) \cite{Neubert}. As for the l.h.s., and considering as intermediate $D$ states the
multiplets $\frac{1}{2}^{-}$, $\frac{1}{2}^{+}$, $\frac{3}{2}^{+}$ (higher
$j^{P}$ intermediate states do not in fact contribute \cite{Jugeau2}), we have
three types of matrix elements coresponding to the hadronic transitions
$B\rightarrow D(\frac{1}{2}^{-})$, $B\rightarrow D(\frac{1}{2}^{+})$ and
$B\rightarrow D(\frac{3}{2}^{+})$. The corrections in $1/m_{b}$ or $1/m_{c}$
to the first matrix element are given by the same ground state subleading IW
functions $L_{i}(w)$, while the $O(1/m_{b})$ and $O(1/m_{c})$ corrections to
the matrix elements $B\rightarrow D(\frac{1}{2}^{+})$, $D(\frac{3}{2}^{+})$ have been carefully
studied \cite{Leibovich} and result in a number of new subleading IW
functions. All these corrections are of two types, perturbations of the heavy
quark current and perturbations of the Lagrangian.

Beginning with the general SR in the heavy quark limit and
perturbing the heavy quark limit matrix elements with $1/m_{b}$ and $1/m_{c}$
corrections, the general expression could then be written, making explicit the
leading and the $1/m_{c}$ and $1/m_{b}$ parts, as :
\begin{equation}
\begin{array}{c}
 
 G_{0}(w_{i},w_{f},w_{if})+E_{0}(w_{i},w_{f},w_{if})+\frac{1}{2m_{b}}[G_{b}(w_{i},w_{f},w_{if})+E_{b}(w_{i},w_{f},w_{if})]\\
\\
+\frac{1}{2m_{c}}[G_{c}(w_{i},w_{f},w_{if})+E_{c}(w_{i},w_{f},w_{if})]
\\
\\
=R_{0}(w_{i},w_{f},w_{if})+\frac{1}{2m_{b}}R_{b}(w_{i},w_{f},w_{if})+\frac{1}{2m_{c}}R_{c}(w_{i},w_{f},w_{if})\label{corr}
\end{array}
\end{equation}
where the subindex 0 means the heavy quark limit, while the subindex $b$ or
$c$ correspond to the subleading corrections in $1/m_{b}$ or $1/m_{c}$, and $G$ or
$E$ mean, respectively, ground state or excited state contributions.

In the heavy quark limit, on has :
\begin{equation}
G_{0}(w_{i},w_{f},w_{if})+E_{0}(w_{i},w_{f},w_{if})=R_{0}(w_{i},w_{f},w_{if})\label{cor}
\end{equation}
that leads to the results quoted in the section 2.

In expression (\ref{corr}), we can vary $m_{b}$ and $m_{c}$ as independent
parameters and obtain new SR for the subleading quantities. To obtain
information on the $1/m_{b}$ corrections, it is relatively simple to proceed as
follows by assuming the formal limit :
\begin{equation}
m_{c}>>m_{b}>>\Lambda_{QCD}
\end{equation}
and perturb both sides of the SR (\ref{cor}) by $1/m_{b}$ terms. In this limit,
one obtains the relation :
\begin{equation}
G_{b}(w_{i},w_{f},w_{if})+E_{b}(w_{i},w_{f},w_{if})=R_{b}(w_{i},w_{f},w_{if}).\label{corbis}
\end{equation}
One can then compute $G_{b}(w_{i},w_{f},w_{if})$ and $E_{b}(w_{i},w_{f},w_{if})$ using
respectively the formalism of Falk and Neubert \cite{Neubert} and the the one of Leibovich
{\it et al.} \cite{Leibovich}, and obtain SR for the different subleading IW functions
$L_{i}(w)$.

Choosing the axial currents $\Gamma_{1}=\not{v_{i}}\gamma_{5}$ and
$\Gamma_{2}=\not{v_{f}}\gamma_{5}$ and taking as initial and final states
respectively the ground state pseudoscalar and vector mesons at different
four-velocities, one obtains the two sum rules \cite{Jugeau2} : 
\begin{equation}
\begin{array}{c}
L_{4}(w)=-6\Sigma_{n}\Delta E_{1/2}^{(n)}\tau_{1/2}^{(n)}(1)\tau_{1/2}^{(n)}(w)\\ 
\\
-L_{5}(w)+(w+1)L_{6}(w)=2\Sigma_{n}\Delta E_{1/2}^{(n)}\tau_{1/2}^{(n)}(1)\tau_{1/2}^{(n)}(w)-4(w+1)\Sigma_{n}\Delta E_{3/2}^{(n)}\tau_{3/2}^{(n)}(1)\tau_{3/2}^{(n)}(w).
\end{array}
\end{equation}
The functions $L_{i}(w)$ ($i=4,5,6$) are not independent but are given in terms of
two functions \cite{Neubert}, namely the elastic IW function $\xi(w)$, a subleading function
$\xi_{3}(w)$ and the fundamental parameter of HQET $\overline{\Lambda}$
($\overline{\Lambda}=m_{B}-m_{b}$, which stands for the energy of the light
cloud). Therefore, we finally obtain the relations, valid for all $w$ :
\begin{equation}
\overline{\Lambda}\xi(w)=2(w+1)\Sigma_{n}^{}\Delta E_{3/2}^{(n)}\tau_{3/2}^{(n)}(1)\tau_{3/2}^{(n)}(w)+2\Sigma_{n}^{}\Delta E_{1/2}^{(n)}\tau_{1/2}^{(n)}(1)\tau_{1/2}^{(n)}(w)\label{newSR1}
\end{equation}
\begin{equation}
\xi_{3}(w)=(w+1)\Sigma_{n}^{}\Delta
E_{3/2}^{(n)}\tau_{3/2}^{(n)}(1)\tau_{3/2}^{(n)}(w)-2\Sigma_{n}^{}\Delta
E_{1/2}^{(n)}\tau_{1/2}^{(n)}(1)\tau_{1/2}^{(n)}(w).
\end{equation}
These new SR reduce to known SR for $w=1$, for
$\overline{\Lambda}$ obtained by Voloshin \cite{Voloshin}, and for $\xi_{3}(1)$ obtained by Le
Yaouanc {\it et al.} \cite{LeYaouanc} and by Uraltsev \cite{Uraltsev}, and
generalize them to all values of $w$. 

Therefore, we have shown that the subleading quantities $\overline{\Lambda}\xi(w)$
and $\xi_{3}(w)$, that are functions of $w$, can be expressed in terms of
leading quantities, namely the transition IW functions $\tau_{j}^{(n)}$ and
the corresponding level spacings $\Delta E_{j}^{(n)}$ $(j=\frac{1}{2}, \frac{3}{2})$.
We can now discuss phenomenological applications of these results,
in particular the check of Bakamjian-Thomas quark models. Within this scheme,
$\xi(w)$ is given by (\ref{IW}) with $\rho^{2}=1.02$, while one gets, for the
$n=0$ states \cite{Morenas} :
\begin{equation}
\tau_{j}^{(0)}(w)=\tau_{j}^{(0)}(1)(\frac{2}{w+1})^{2\sigma_{j}^{2}}
\end{equation}
with $\tau_{1/2}^{(0)}(1)=0.22$, $\sigma_{1/2}^{2}=0.83$, $\tau_{3/2}^{(0)}(1)=0.54$
and $\sigma_{3/2}^{2}=1.50$. Assuming the reasonable saturation of the SR with
the lowest $n=0$ states, one gets, from the first relation (\ref{newSR1}), a
sensibly constant value for $\overline{\Lambda}=0.513\pm0.015$.

\section{Conclusion}

Using sum rules in the heavy quark limit of QCD, as formulated in ref. \cite{Oliver}, lower bounds for the derivatives of the elastic
IW function $\xi(w)$ have been found. Any phenomenological parametrization of $\xi(w)$
intending to fit the CKM matrix element $|V_{cb}|$ in $B\rightarrow
D^{(\ast)}l\nu$ should satisfy these bounds. Moreover, we have found non-trivial new information on
subleading contributions in $1/m_{Q}$. As a result, the fundamental quantity
of HQET $\overline{\Lambda}$ appears to be a ratio of leading functions and
$\xi_{3}(w)$ is also given in terms of leading quantities. To proceed further
phenomenologically, we use as an Ansatz for these functions the results of the
Bakamjian-Thomas quark models, that gives covariant form factors in the heavy
quark limit, satisfies IW scaling and also Bjorken and Uraltsev sum rules. One
obtains in this way for a very wide range of $w$ the expected constancy for
$\overline{\Lambda}$, with numerical value of the order of 0.5 and which is in
agreement with other methods as, for example, the QCDSR approach. 

\section*{Acknowledgments}
I am indebted to B. Blossier, J. Hirn, G. Therin and J. Welzel for very useful
discussions. I would like also to thank the
organizators who gave me the opportunity to present my research activities during this conference. I acknowledge support from the program ``Human resources and
Mobility-Marie Curie Conferences''.

\end{document}